\documentclass[fleqn,usenatbib]{mnras}

\usepackage{newtxtext,newtxmath}

\usepackage[T1]{fontenc}

\DeclareRobustCommand{\VAN}[3]{#2}
\let\VANthebibliography\thebibliography
\def\thebibliography{\DeclareRobustCommand{\VAN}[3]{##3}\VANthebibliography}

\usepackage{graphicx}	
\usepackage{amsmath}	
\usepackage{xcolor}

\title[Warps on barred and non-barred galaxies]{Warps induced by satellites on barred and non-barred galaxies}

\author[A. Wille \& R. E. G. Machado]{
A. Wille$^{1}$\thanks{E-mail: andressawille@alunos.utfpr.edu.br} and
R. E. G. Machado$^{1}$
\\
$^{1}$Departamento Acadêmico de Física, Universidade Tecnológica Federal do Paraná, Av. Sete de Setembro 3165, 80230-901 Curitiba, Brazil\\
}

\date{Accepted 2024 August 16. Received 2024 August 2; in original form 2024 May 8}

\pubyear{2024}

\begin{document}
\label{firstpage}
\pagerange{\pageref{firstpage}--\pageref{lastpage}}
\maketitle

\begin{abstract}
Warps are common vertical asymmetries that appear in the outer parts of the galactic discs, bending one part upwards and the other downwards. Many mechanisms can trigger warp formation, including tidal interactions. The interactions with satellites distort the edges of the disc and can also change the central morphology, impacting, for example, the development of a galactic bar. In mergers events, the bar can be weakened or even destroyed. In this study, we aim to compare barred and non-barred galaxy models and their susceptibility to warping. To analyze the effects of induced warps, we used $N$-body simulations of a barred and a non-barred central galaxy interacting with satellites of varying masses ($0.1 \times 10^{10} \mathrm{M_{\odot}}$, $0.5 \times 10^{10} \mathrm{M_{\odot}}$ and $1 \times 10^{10} \mathrm{M_{\odot}}$) and initial orbital radii (10, 20 and 30 kpc). We also ran isolated simulations of the central galaxies for comparison. We found that the induced warps are stronger in the barred galaxy compared with the non-barred galaxy, in perturbed and isolated models. In addition, the masses of the satellites determine the level of destruction of the bar and the intensity of the induced warp. The time in which the bar will be weakened or destroyed depends on the orbital radius of the satellite.

\end{abstract}





\section{Introduction}
\label{sec:intro}

Morphological asymmetries, such as rings and wave-like structures, are common in spiral galaxies. Warps are vertical asymmetries in the galactic disc, where the amplitude increases with radius. The most common kind of distortion is the S-shaped or “integral sign” warp, but U-shaped warps or more complex structures can also appear \citep{Gomez2017}.

Detecting warps is not a trivial task, since they can only be seen when observing edge-on galaxies, and even with this orientation, not all angles are ideal for detection and measurements. In addition, tilted structures can be contaminated by dust or project spiral arms \citep{RC1998}. Surveys of HI 21 cm detected warp signals in several galaxies in the 1970s \citep{binney1992}, and since then, many observational studies have been carried out, especially on the warp of the Milky Way, already known at the time \citep{burke57}. Statistical studies have shown how common this type of structure is, reaching up to 70 per cent of the observational samples analyzed \citep[e.g.][]{RC1998, ann}.

With more robust mapping surveys, it is possible to perform more detailed investigations into the structure of our galaxy. Recent studies have provided more insights into the aspects of the Galactic warp based on observational data \citep[e.g.][]{Huang_2018,Schonrich2018,romero2019,cheng, dehnen2023} and also on simulations \citep[e.g.][]{laporte,Poggio2021,bland_tepper}. Many of these works point out that the interaction with the Sagittarius dwarf galaxy probably causes the bending waves in the Milky Way.

Because of the challenges of observational studies on warps, simulations can be a very useful instrument to understand the evolution of vertical perturbations and measure their characteristics, such as duration, amplitude and, most importantly, the mechanisms that can trigger warp formation. Some mechanisms are accretion of misaligned cold gas in the galactic disc \citep{Gomez2017,k2022}, misalignment between the stellar disc and the inner dark matter halo \citep{han2023, garcia-conde}, and even more subtle factors, such as misalignment between the angular momentum and the angular velocity in models of barred galaxies \citep{sanchez2016} or random noise in the distribution of halo and bulge particles in $N$-body simulations \citep{chequers}. In addition, the role of external perturbations (fly-bys and interactions with satellites) is reinforced by many studies \citep[e.g.][]{Gomez2017,semczuk,tepper2022,garcia-conde}. Despite all the other possibilities, tidal interactions seem to be the most obvious mechanism that induces features in the outer disc.

In addition to vertical asymmetries, spiral galaxies can have other features, such as bars. Nearly 60 per cent of spiral galaxies are barred \citep{eskridge2000,sheth2008,lee2019} when observed in near infrared. Bars are elongated stellar structures in the center of the galactic disc, and they can have different masses, shapes, and kinematics \citep{misiriotis2002}. Bar formation can occur in isolation, via the exchange of angular momentum between the inner disc and the halo or outer disc: circular orbits in the center of the galaxy become more elongated, losing angular momentum and forming the bar \citep{athanassoula2003}. But bars can also be induced by tidal interactions, especially if the perturber is on a prograde orbit is massive or has a close orbit \citep{lokas2018, peschken}.  In simulations, strong bars are preferentially formed in galaxies with spherical halos and with little or no gas \citep{AMR2013}.

The evolution of galaxies is directly linked to the evolution of bars. Only a sufficiently massive and cold disc can form a bar \citep{sheth2008}, and once the bar starts to grow, it has a direct impact on the evolution of disc parameters, such as density profiles \citep{debattista}. However, the study of these structures has limitations and difficulties: optical observations show a smaller fraction of barred galaxies than infrared observations \citep{eskridge2000} and the orientation of the galaxy can make it difficult to identify the bar, which is best seen in a face-on objects. The Milky Way also has a bar that is difficult to observe due to our position in the galaxy and the dust extinction \citep{shen2020}. Because of that, numerical simulations are a valuable tool in galactic and extragalactic astronomy.

In $N$-body simulations, the bars form spontaneously in a few Gyr, but can also be weakened or destroyed. Bar dissolution can occur because of central mass concentrations, such as supermassive black holes, large condensations of molecular gas and star clusters \citep[e.g.][]{das2003,shen_sellwood,hozumi}, or if the galaxy is in a prolate halo \citep{ideta2000}. Furthermore, interactions with satellites can cause bar weakening in the case of minor merger events \citep{ghosh} or bar destruction in dense environments \citep{rosas2024}, and if massive satellites collide with the disc the bar is also destroyed \citep{athanassoula2003p}. However, bars can be resilient: unequal-mass encounters can weaken the bar, but it is recovered after the interaction \citep{zana}.

There are still not many papers that specifically explore the formation of warps in barred galaxies. In this study, we aim to understand if a barred galaxy is more resilient against warping, or more susceptible to it, compared to a galaxy that does not have a bar. We use $N$-body simulations of galaxies interacting with satellites of varying masses and initial orbital radii to analyze the effects of warps induced by these satellites on barred and non-barred galaxies. This paper is organized as follows: in section \ref{sec:methods}, we describe the details of the simulation setup and the characteristics of central galaxies and satellites; section \ref{sec:results} presents the results of the interactions and analysis about bar aspects and warp development; section \ref{sec:discussion} includes the discussion and section \ref{sec:conclusions} contains the conclusions.


\section{Simulations setup}
\label{sec:methods}

In this section, we describe the properties of the initial conditions of the simulations. The initial conditions are composed of a massive central galaxy (which can be barred or non-barred), and a satellite galaxy orbiting it. In the first part of this section, we describe the central galaxy models and what parameters make their morphologies distinct. Next, we describe the satellites, which can have different masses and orbital radii.


\subsection{Central galaxies}
\label{sec:central-galaxies}

Our first central galaxy model is a spiral galaxy with a strong bar (model B). It is composed of a disc, a bulge and a dark matter halo. The disc density profile is given by equation~(\ref{eq:disc}):

\begin{equation}
\rho(R, z) = \frac{M_\mathrm{d}}{4\pi R_0^2 z_0} \, \mathrm{exp} \left(\frac{-R}{R_0}\right) \, \mathrm{sech}^2 \left(\frac{z}{z_0}\right),
 \label{eq:disc}
\end{equation}
where the disc mass is $M_\mathrm{d} = 4 \times 10^{10} \mathrm{M_{\odot}}$, the radial scale length is $R_0=3.5$ kpc and the vertical scale length is $z_0 = 0.7$ kpc. For the halo, we used the \cite{hernquist} profile, given by equation~(\ref{eq:halo}):

\begin{equation}
\rho(r) = \frac{M_\mathrm{h}}{2\pi} \, \frac{a}{r} \, \frac{1}{(r+a)^3} \,,
 \label{eq:halo}
\end{equation}
where the scale length is $a=47$ kpc and the halo mass is $M_\mathrm{h} = 1 \times 10^{12} \mathrm{M_{\odot}}$. Finally, the bulge follows the same density profile, but with a smaller scale length, $a =1.5$ kpc. It has a mass of $M_\mathrm{b} = 1 \times 10^{10} \mathrm{M_{\odot}}$, i.e. it corresponds to 20 per cent of the total baryonic mass.

We used the galstep code \citep{ruggiero} to create the initial conditions, setting the masses mentioned above and an appropriate number of particles for each component. The disc has $N_\mathrm{d}=2 \times 10^{5}$ particles, the bulge has $N_\mathrm{b}=0.5 \times 10^{5}$ particles, and the halo has $N_\mathrm{h}=1 \times 10^{6}$ particles. After establishing the initial conditions, we used the Gadget-4 code \citep{springel} to run the simulations, using a gravitational softening length of $\epsilon = 0.1$ kpc for all the components. 

In the analysis of galactic bars in $N$-body simulations, it is usually sufficient for the galaxy to evolve for a few gigayears. However, in this study, we need to extend the simulation longer. The introduction of satellites must happen when the bar in the central galaxy is already well-developed and robust, and then, the interaction between the two galaxies must last a few more Gyr to be effective. For this reason, we let the model evolve for a total 14 Gyr, saving outputs every 0.05 Gyr.

\begin{figure*}
    \includegraphics[width=2\columnwidth]{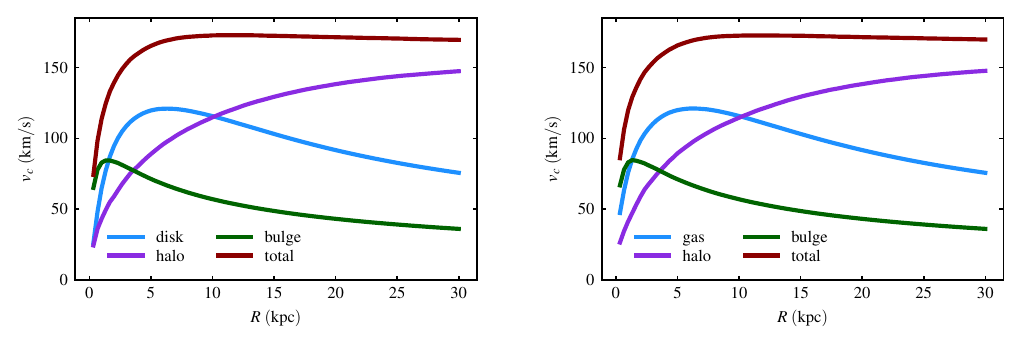}
    \caption{Left: circular velocity curve of the components of the barred galaxy. Right: circular velocity curve of the components of the non-barred galaxy.}
    \label{fig:fig1-nb}
\end{figure*}

For comparison in future analyses, we need a second model, a galaxy with the same size and mass, but non-barred. Bars appear spontaneously in a large number of studies with $N$-body simulations \citep{misiriotis2002}, but the presence of gas in the disc is known to inhibit the formation of strong bars in simulated galaxies \citep{AMR2013}. Then, this model is a disc galaxy with a weak bar, but because of the contrast with the much stronger bar in model B, it is reasonable to label it non-barred (model NB). It has gas, bulge and dark matter halo. The initial conditions were also created with the galstep code, but some parameters have changed. Disc particles were entirely replaced by gas particles. The gas also has an exponential density profile described by equation~(\ref{eq:disc}), but with $z_0 = 0.035$ kpc. The other components (halo and bulge) follow the same density profiles shown in equations~(\ref{eq:halo}). The halo mass is also $M_\mathrm{h} = 1 \,\times \, 10^{12} \mathrm{M_{\odot}}$ and the baryonic mass is still $5 \, \times \,10^{10} \mathrm{M_{\odot}}$, distributed between gaseous disc and bulge. The gas mass is $M_\mathrm{g} = 4 \times 10^{10} \mathrm{M_{\odot}}$ and the bulge mass is $M_\mathrm{b} = 1 \times 10^{10} \mathrm{M_{\odot}}$. This model includes cooling and star formation. We used the same gravitational softening length and the outputs were also saved every 0.05 Gyr. Fig. \ref{fig:fig1-nb} compares the circular velocity curves of models B and NB. The only difference in the initial conditions is that B has a fully stellar disc, while NB has a fully gaseous disc.

\begin{figure}
    \includegraphics[width=\columnwidth]{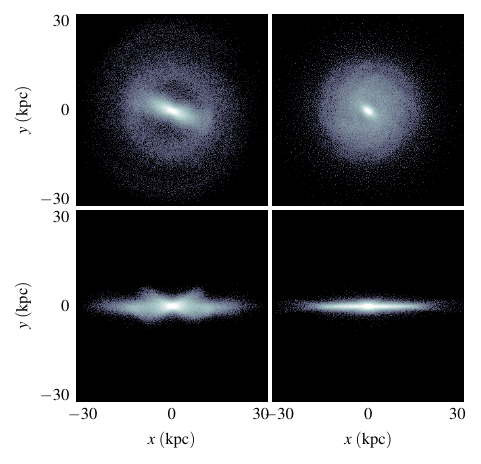}
	\caption{Morphology of the galaxies at the end of the simulation ($14$ Gyr). On the left, the barred galaxy (model B), seen face-on in the top panel and edge-on in the bottom panel. On the right, the non-barred galaxy (model NB), also seen in the face-on and edge-on projections.}
	\label{fig:fig2}
\end{figure}

At the end of the simulations, the presence or absence of gas in the galactic disc resulted in models B and NB having very different morphologies. Fig.~\ref{fig:fig2} shows the final face-on and edge-on projections of the galaxies. Our first model ended up being a strongly barred galaxy with a peanut-shaped bar, and the second model is a galaxy that does not develop a substantial bar, maintaining a more axisymmetric shape. The bar in NB model is not only weaker, but also much shorter.

In order to quantify this difference in the central part of the galaxies, we calculated the bar strength $A_2$ for each one. We used the maximum value of the relative amplitude of the $m=2$ mode of the projected mass distribution \citep{AMR2013, lokas2019, peschken,cuomo}. This mode is sensitive to non-axisymmetric configurations, which is the case for an elongated structure like the bar. $A_2$ is given by equation~(\ref{eq:a2}):

\begin{equation}
A_2 = \mathrm{max} \left(\frac{\sqrt{a_2^2+b_2^2}}{a_0}\right),
 \label{eq:a2}
\end{equation}
where $a_{m}$ and $b_{m}$ are the Fourier coefficients shown by equations~(\ref{eq:a}) and (\ref{eq:b}):

\begin{equation}
a_{m} (R) = \sum_{i=0}^{N_R} m_i\, \mathrm{cos}(m \, \theta_i), \, m = 0,1,2...
 \label{eq:a}
\end{equation}

\begin{equation}
b_{m} (R) = \sum_{i=0}^{N_R} m_i\, \mathrm{sen}(m  \,\theta_i), \, m = 1,2...
 \label{eq:b}
\end{equation}
with $m_i$ being the mass in each ring $i$ of radius $R$.

\begin{figure}
    \includegraphics[width=\columnwidth]{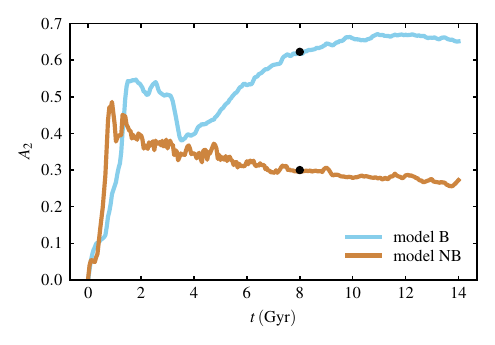}
	\caption{Evolution of bar strength $A_2$. The black points mark the moment when the satellite galaxies are added to the simulation, at $t=8$ Gyr.}
	\label{fig:fig3}
\end{figure}

Fig.~\ref{fig:fig3} shows the time evolution of $A_2$ for the two models. In both cases, the full baryonic mass of the disc was used used to compute this quantity, as well as the subsequent measurements. As expected, the barred galaxy has a much higher final $A_2$ value than the non-barred galaxy. It has four clear phases: rapid growth, an almost flat phase, a sudden decrease (buckling), and a smooth secular evolution. Despite ending up with low $A_2$ values, the non-barred galaxy shows some peaks at the beginning of the simulation, but they become smoother with time. In Fig.~\ref{fig:fig3}, it also can be noted that the bar is already very robust in model B and that there are no instabilities in model NB at $t=8$ Gyr, the time chosen for the satellite galaxies to be introduced into the simulations to orbit the central galaxies.

\subsection{Satellite galaxies}
\label{sec:satellites}

The initial conditions of the satellites were created with the Hernquist density profile defined in equation~(\ref{eq:halo}). The scale length is $a=0.4$ kpc and the masses are $0.1 \times 10^{10} \mathrm{M_{\odot}}$, $0.5 \times 10^{10} \mathrm{M_{\odot}}$ and $1 \times 10^{10} \mathrm{M_{\odot}}$, which represents, respectively, 2, 10 and 20 per cent of the baryonic mass of the central galaxy that the satellites will orbit. These three masses were labeled as M1, M2 and M3. The effect of the satellites on the host galaxy depends mainly on their mass, not on their internal structure. Since dark matter is the predominant component of the total mass, we created the initial conditions with dark matter only. We have verified that satellites built this way are stable if they evolve in isolation.

\begin{figure}
    \includegraphics[width=\columnwidth]{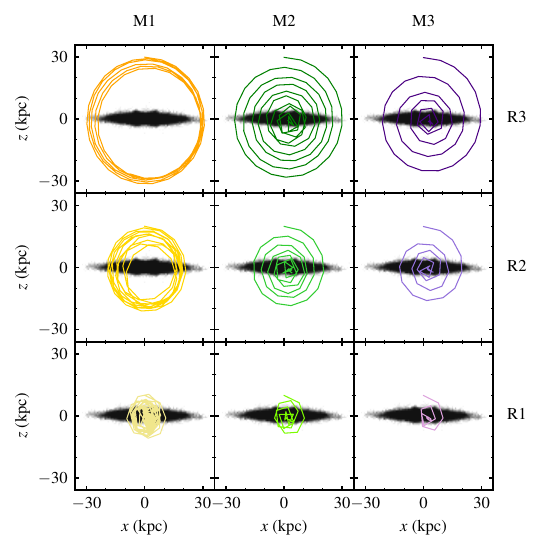}
	\caption{Satellite orbits around the barred galaxy during 6 Gyr (from 8 to 14 Gyr). From left to right, the masses of the satellites increase ($0.1 \times 10^{10} \mathrm{M_{\odot}}$ (M1), $0.5 \times 10^{10} \mathrm{M_{\odot}}$ (M2) and $1 \times 10^{10} \mathrm{M_{\odot}}$ (M3)), and from top to bottom, the orbital radii decrease ($30$ kpc (R3), $20$ kpc (R2) and $10$ kpc (R1)). Thus, the yellowish tones characterize the lower mass satellites, the green tones characterize the intermediate mass satellites, and the purple tones the higher mass satellites. Light shades represent smaller radii, and darker shades represent larger radii.}
	\label{fig:fig4}
\end{figure}

\begin{figure}
    \includegraphics[width=\columnwidth]{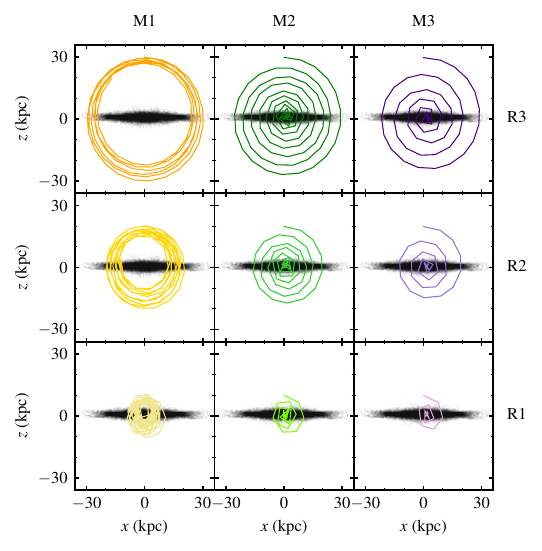}
	\caption{Satellite orbits around the non-barred galaxy during 6 Gyr. The color scheme is the same as in Fig.~\ref{fig:fig4}.}
	\label{fig:fig5}
\end{figure}

The satellite galaxies were positioned in polar orbits around to the central galaxy, with initial radii of $10$ kpc, $20$ kpc and $30$ kpc. These radii were labeled as R1, R2 and R3. All the orbits are initially circular and last from 8 to 14 Gyr. The initial velocities of the satellites are the circular velocities corresponding to each radius: 165 km\,s$^{-1}$, 171 km\,s$^{-1}$ and 169 km\,s$^{-1}$ for radii R1, R2 and R3, respectively. As the rotation curve of the central galaxies is approximately flat in the 10 -- 30 kpc radial range, it is natural that these velocities are similar to each other. Figs.~\ref{fig:fig4} and \ref{fig:fig5} show each of the 9 models around galaxy B, and then again around galaxy NB. Note that in all cases, the satellite galaxy ends up absorbed, except in models M1R3 and M1R2, where the simulation ends without merging. At each time step, the position of the satellite galaxy was determined using the density peak of its particles. In cases where the satellite has been absorbed by the central galaxy, this peak is not well defined.

Since the central galaxies can have different morphologies and the satellites can have different masses and orbital radii, Table~\ref{tab:tab1} summarizes the main information about these variables and identifies each model. For each of the 3 masses of the satellite galaxy, we have 3 orbital radii. Therefore, we have 9 models around galaxy B, and another 9 models around galaxy NB, resulting in a total of 18 simulations. There are also the models B and NB that have been continued in isolation. Each one will have the central part and edges of the disc affected differently, depending on the combination of the parameters.

\begin{table}
	\centering
	\caption{Names and parameters for each model.}
	\label{tab:tab1}
	\begin{tabular}{cccc} 
		\hline
		Model name & Central galaxy & Satellite mass & Satellite initial \\
         &   &   ($10^{10} \mathrm{M_{\odot}}$) &  radius (kpc) \\
		\hline
        B-M1R1 & Barred & $0.1$ & $10$ \\
        B-M2R1 & Barred & $0.5$ & 10 \\
        B-M3R1 & Barred & $1$ & 10 \\[0.5em]
        B-M1R2 & Barred & $0.1$ & 20 \\
        B-M2R2 & Barred & $0.5$ & 20 \\
        B-M3R2 & Barred & $1$ & 20 \\[0.5em]
        B-M1R3 & Barred & $0.1$ & 30 \\
        B-M2R3 & Barred & $0.5$ & 30 \\
        B-M3R3 & Barred & $1$ & 30 \\[0.5em]
        NB-M1R1 & Non-barred & $0.1$ & 10 \\
        NB-M2R1 & Non-barred & $0.5$ & 10 \\
        NB-M3R1 & Non-barred & $1$ & 10 \\[0.5em]
        NB-M1R2 & Non-barred & $0.1$ & 20 \\
        NB-M2R2 & Non-barred & $0.5$ & 20 \\
        NB-M3R2 & Non-barred & $1$ & 20 \\[0.5em]
        NB-M1R3 & Non-barred & $0.1$ & 30 \\
        NB-M2R3 & Non-barred & $0.5$ & 30 \\
        NB-M3R3 & Non-barred & $1$ & 30 \\
		\hline
	\end{tabular}
\end{table}


\section{Results}
\label{sec:results}
In this section, we show the main results of our investigations. We first focus on the analysis of how the different satellite models affect the bar development in the barred galaxy. In the second part, we show the influence of the satellite on the occurrence of warps and we compare the intensity of this phenomenon in the barred and non-barred galaxies.


\subsection{How the pertuber affects bar strength}
\label{sec:bar}

\begin{figure}
    \includegraphics[width=\columnwidth]{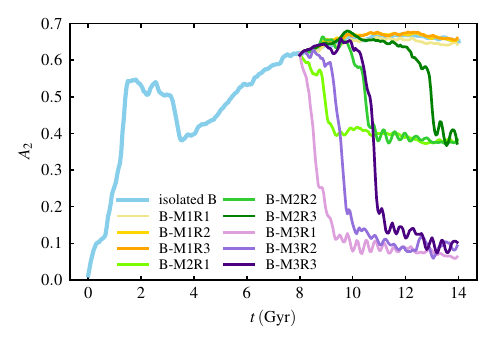}
	\caption{Evolution of bar strength in isolated and perturbed models of the barred galaxy. Again, the color scheme is the same as in Fig.~\ref{fig:fig4}.}
	\label{fig:fig6}
\end{figure}

The close passage of the satellites caused changes in the strength and the shape of the bar in model B. To quantify these changes, we calculated $A_2$ for all the perturbed models to compare it with the isolated ones. In Fig.~\ref{fig:fig6} it is evident that bar strength decrease is directly related to the mass and orbit of the satellite. The mass influences in how much the bar will lose strength: less massive perturbers (M1) barely affect $A_2$, and the bar remain as strong as in the isolated B model; satellites of intermediate mass (M2) weaken the bar without destroying it; and the most massive satellites (M3) destroy the bar, as evidenced by the large decline in $A_2$ values. In addition to the mass, another parameter of the perturber is relevant: the satellite’s orbital radius determines the moment where strength decrease will occur. For satellites with masses M2 and M3, it is noticeable in Fig.~\ref{fig:fig6} that the closer the satellite's orbit, the sooner the bar is attenuated or destroyed, compared to more distant orbits. Therefore, the changes in $A_2$ occur first for satellites with initial orbital radius R1, then R2 and finally R3. We did the same analysis for the non-barred galaxy (not shown). For all its interactions, the final $A_2$ value remains low as in the isolated model NB. 

\begin{figure}
    \includegraphics[width=\columnwidth]{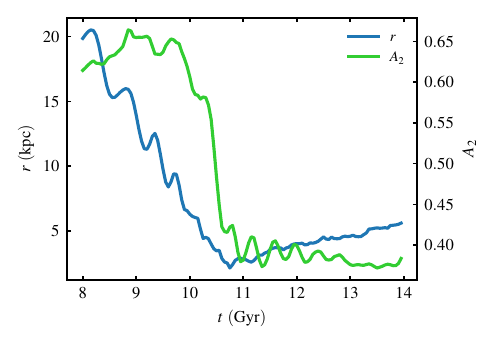}
	\caption{Orbital radius and $A_2$ evolution for model B-M2R2. Note the different axes for each variable. As the satellite approaches the center of the galaxy, the bar strength decreases.}
	\label{fig:fig7}
\end{figure}

Since the initial radius of the orbit determines whether the decline in bar strength will happen earlier or later, we also analyzed the evolution of the satellites orbital radius $r$ together with the evolution of the bar strength in the central galaxy. Fig.~\ref{fig:fig7} shows one example (for model B-M2R2) of the temporal correlation between the greater proximity of the satellite and lower $A_2$ values. This result applies to all other simulations with intermediate (M2) and massive (M3) satellites as well. 

\begin{figure*}
    \includegraphics[width=2\columnwidth]{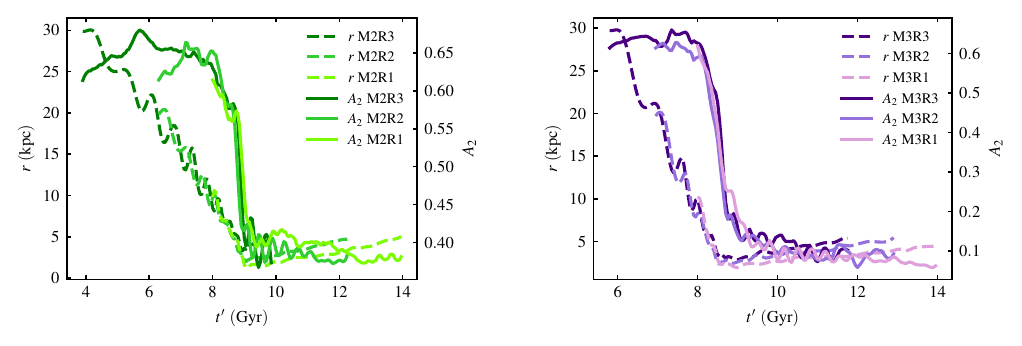}
	\caption{Orbital radius and $A_2$ evolution for models perturbed by M2 (left) and M3 (right) satellites. A shift in time causes an alignment of the curves. When the perturber reaches a radius of around 5 kpc, there is a drop in $A_2$, indicating that the bar has been weakened or destroyed.}
	\label{fig:fig7-1}
\end{figure*}

Additionally, to identify if this significant decrease in $A_2$ occurs when the satellite reaches a particular radius $r$, we compared the evolution of the orbital radius and the bar strength, shifting the time arbitrarily until the $r$ curves (dashed) aligned. This resulted in the $A_2$ curves (solid) also aligning, as shown in Fig.~\ref{fig:fig7-1}. The models with satellite mass M2 are on the left and those with satellite mass M3 are on the right. On the horizontal axis, $t'$ is not the physical time at which all the $A_2$ declines occur (they occur at different times, as already mentioned about Fig.~\ref{fig:fig6}), but a shifted time. In the case of the M2 satellites, the time $t$ of the M2R1 model was kept fixed, and a shift of $\Delta t = -1.7$ Gyr and $\Delta t = -4.1$ Gyr was applied, respectively for M2R2 and M2R3 models. These values were enough for the curves to align. The same was done for the case of the M3 satellites, but with $\Delta t = -1.05$ Gyr and $\Delta t = -2.2$ Gyr applied to M3R2 and M3R3 curves. The aligned curves indicate that when the satellite reaches an orbital radius of $\sim 5$ kpc there is a sudden drop in the bar strength.

\subsection{Development of the warps}
\label{sec:warps}

\begin{figure}
    \includegraphics[width=\columnwidth]{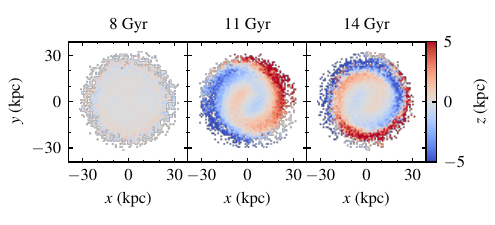}
	\caption{Mean heights map for model B-M3R1 at different times, showing the difference in the heights $z$ of the edge of the galactic disc after interaction with the satellite.}
	\label{fig:fig8}
\end{figure}

The formation of warps has also been noted during the interaction between the central galaxies and the satellites. We initially studied maps of mean heights $z$ in order to qualitatively characterize the differences between the models. Fig.~\ref{fig:fig8} shows an example of a mean heights map for model B-M3R1 at 3 selected times. The bar has been rotated to coincide with the $x$-axis. At $t=8$ Gyr, the disc of the central galaxy has mean heights close to zero, indicating that the disc is initially symmetric with respect to the $z=0$ plane, but within a few Gyr, the edges of the disc are deformed by the gravitational influence of the satellite, and one part is bent upwards and the other downwards, characterizing the S-shaped warp. This morphology is preserved until the end of the simulation. The process is similar for the other models.

\begin{figure}
    \includegraphics[width=\columnwidth]{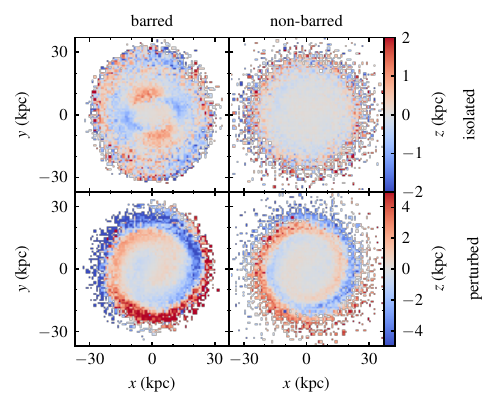}
	\caption{Top panel: map of mean heights for models B and NB (isolated) at $t=14$ Gyr. In model B, the bar can be seen in the center. Bottom panel: map of mean heights for models B-M3R1 and NB-M3R1 at $t=14$ Gyr. Note the different ranges in the colorbars.}
	\label{fig:fig9}
\end{figure}

\begin{figure*}
    \includegraphics[width=2\columnwidth]{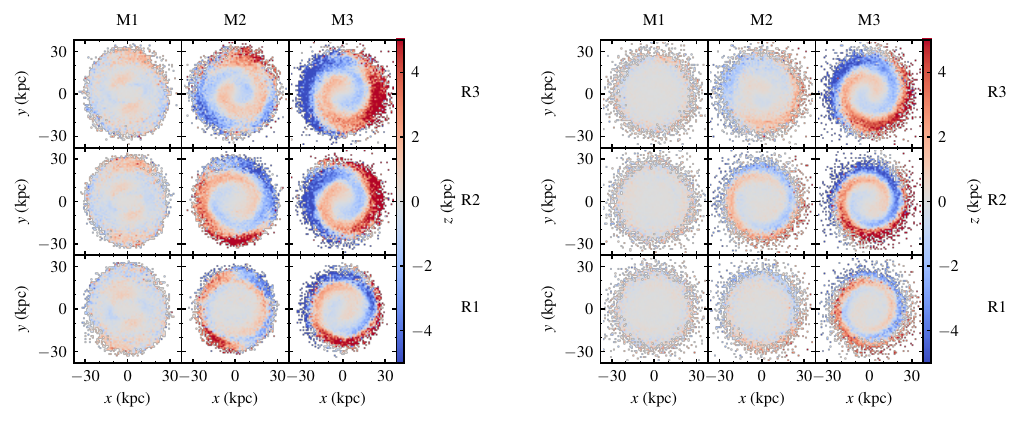}
	\caption{Map of mean heights at $t=14$ Gyr for all models of perturbed barred galaxy (left) and non-barred galaxy (right). As in Fig.~\ref{fig:fig4}, from left to right, the masses of the satellites increase, and from top to bottom, the orbital radii decrease.}
	\label{fig:fig10}
\end{figure*}

When comparing barred and non-barred galaxies, we noted differences in the intensities of the warps induced by the same satellite. Fig.~\ref{fig:fig9} presents the morphologies of the galaxies at $t=14$ Gyr considering the isolated cases and the interaction with a satellite of mass $1 \times 10^{10} \mathrm{M_{\odot}}$ and orbital radius 10 kpc. For the barred galaxy, the bar has been rotated to coincide with the $x$-axis, but this is only noticeable in the top panel, where the range of $z$ is smaller. In the bottom panel, the range is larger, and at the time shown, the bar has already been destroyed by interaction with the most massive satellite. There is no clear warp signal in the isolated cases (top panel), but in the perturbed ones (bottom panel), the barred galaxy exhibits a more intense warp than the non-barred galaxy. This pattern was also found in all the other models, as shown in Fig.~\ref{fig:fig10}.

Fig.~\ref{fig:fig10} show the mean heights maps for barred galaxy (left) and non-barred galaxy (right). As in figures above, the bar has been rotated in the B models. Comparing each NB panel with its equivalent in B, it is clear that the intensity of the induced warp is greater in the barred galaxy in each case, even in cases where the satellite does not merge with the central galaxy (models M1R2 and M1R3). There is also an indication that the intensity depends essentially on the mass of the perturber, for both B and NB galaxies. The lowest mass satellites (M1) cause very small asymmetries in the outer disc of the central galaxies, and the larger mass satellites (M2 and M3) induce more pronounced warps, with the strongest ones resulting from from the perturbation of satellites with mass M3. So for each orbital radius R, the warp intensity increases systematically with the mass M. On the other hand, for a given M, the warp intensity is not clearly correlated with R.

\begin{figure}
    \includegraphics[width=\columnwidth]{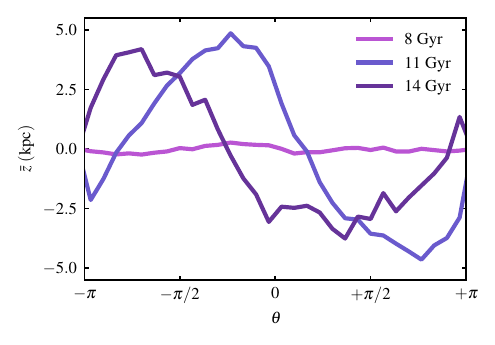}
	\caption{Mean heights as a function of the azimuthal angle $\theta$ for model B-M3R1 at three selected times.}
	\label{fig:fig11}
\end{figure}

\begin{figure}
    \includegraphics[width=\columnwidth]{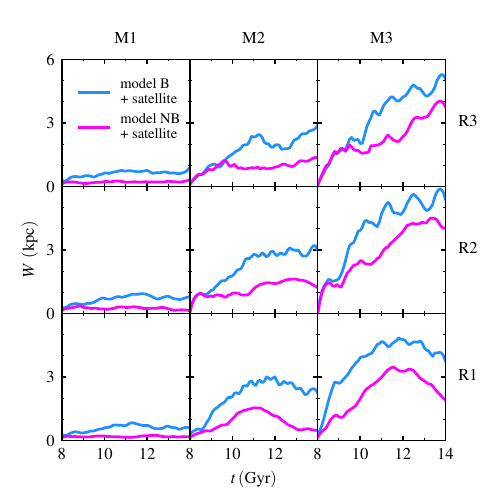}
	\caption{Evolution of maximum amplitude of warp $W$ for all perturbed models.}
	\label{fig:fig12}
\end{figure}

Once these differences between B and NB models were noted, we performed a quantitative analysis of the warp intensity using two methods. In the first, we considered the time evolution of heights $z$ 
as a function of the azimuthal angle $\theta$ in the $xy$ plane. This analysis was carried out in an outer ring of the disc ($20 < R < 30$ kpc, where $R$ is the cylindrical radius). Fig.~\ref{fig:fig11} shows an example for model B-M3R1 at 3 selected times (same model and selected times as Fig.~\ref{fig:fig8}). At each instant of time, there is a maximum and a minimum value of $z$, which characterize, respectively, the part of the edge of the disc that is tilted upwards and the part that is tilted downwards. The maximum amplitude of the warp $W$ is defined by equation (\ref{eq:w}):

\begin{equation}
W (t) = \frac{z_{\mathrm{max}} + |z_{\mathrm{min}}|}{2}.
 \label{eq:w}
\end{equation}
The evolution of $W$ is shown in Fig.~\ref{fig:fig12} for the interacting models. To ensure that the results of this measurement do not depend strongly on the choice of ring, we also tested with a more internal ring ($10 < R < 20$ kpc), which is not shown here, but displays the same pattern. Fig.~\ref{fig:fig12} supports what the mean height maps indicated: the warp is more intense in the barred galaxy in all cases, regardless of the perturbation, and the amplitude of the warp depends on the mass of the perturber. The orbital radius of the satellite does not seem to be a decisive factor.

\begin{figure}
    \includegraphics[width=\columnwidth]{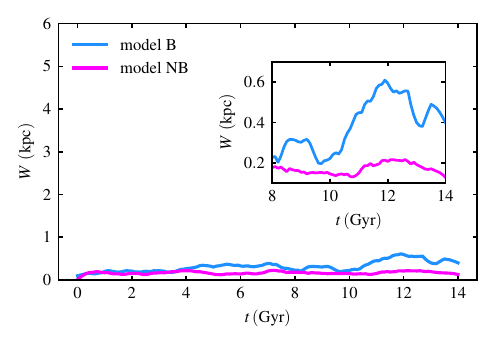}
	\caption{Evolution of maximum amplitude of warp $W$ for isolated models B and NB.}
	\label{fig:fig13}
\end{figure}

If there is such a difference of intensity in the perturbed models, it is reasonable to question if the barred galaxy would have an intrinsic tendency to develop more intense warps even in isolation. In Fig.~\ref{fig:fig13}, we have the evolution of $W$ for the isolated models. We used the same amplitude range as in Fig.~\ref{fig:fig12} for comparison. During most of the simulation, there is almost no disparity in the values of $W$ in the B and NB galaxies, but at approximately 10 Gyr, a small warp signal in the barred galaxy appears without any disturbance. The inset in Fig.~\ref{fig:fig13} highlights this increase in $W$ in the barred galaxy, while the value remains almost constant in the non-barred galaxy.

\begin{figure*}
    \includegraphics[width=2\columnwidth]{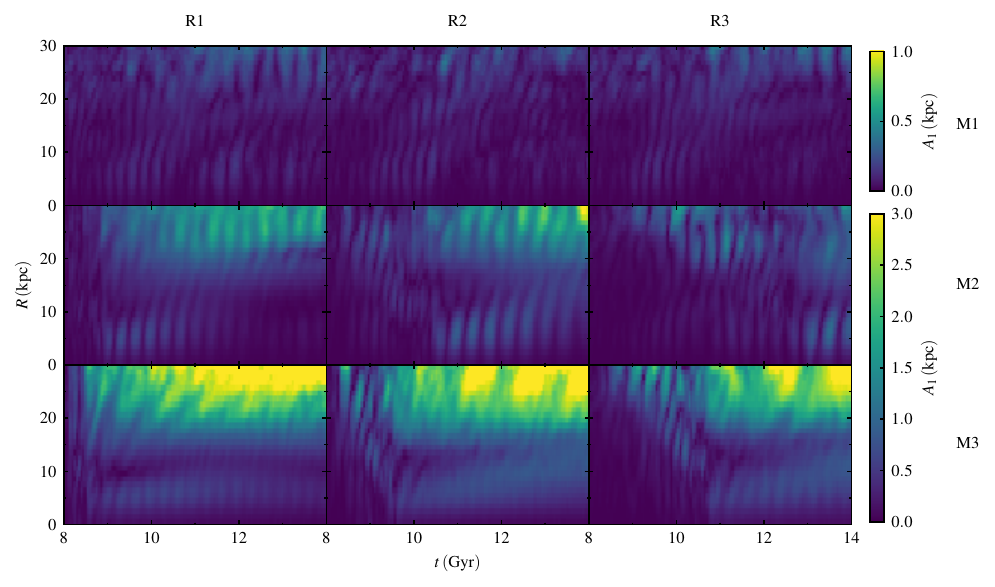}
	\caption{$A_1$ for all perturbed models of the barred galaxy. Note the different ranges in the colorbars.}
	\label{fig:fig14}
\end{figure*}

\begin{figure*}
    \includegraphics[width=2\columnwidth]{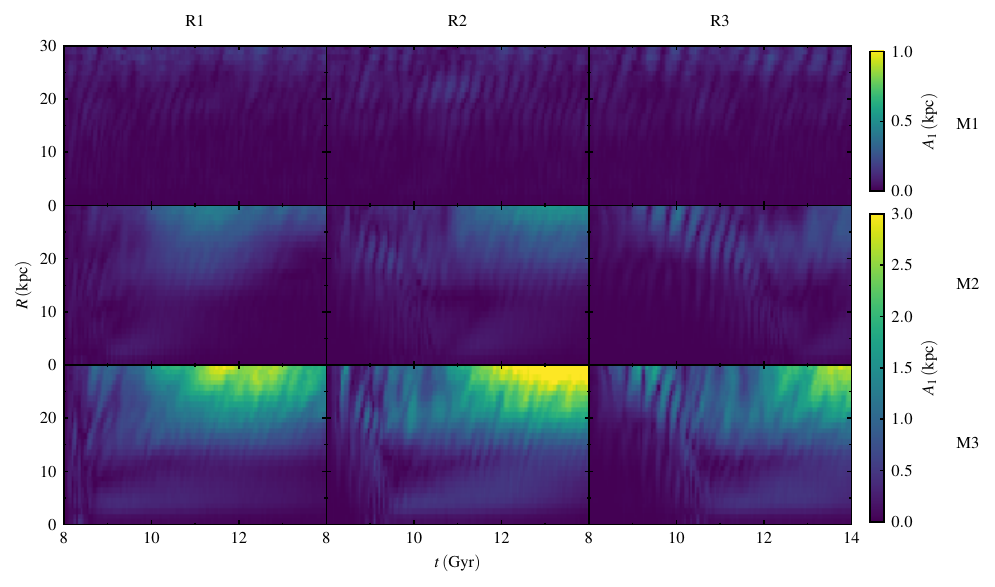}
	\caption{$A_1$ for all perturbed models of the non-barred galaxy. Note the different ranges in the colorbars.}
	\label{fig:fig15}
\end{figure*}

The second method for measuring the vertical distortions is calculating $A_1$, using Fourier components $a_1$ and $b_1$, corresponding to the $m=1$ mode, according to the equations (\ref{eq:a1}) and (\ref{eq:b1}):
\begin{equation}
a_m (R) =   \frac{1}{N_R} \sum_{i=0}^{N_R} z_i\, \mathrm{cos}(m \, \theta_i), \, m = 0,1,2...
 \label{eq:a1}
\end{equation}

\begin{equation}
b_m (R) =  \frac{1}{N_R} \sum_{i=0}^{N_R} z_i\, \mathrm{sen}(m \,\theta_i), \, m = 1,2...
 \label{eq:b1}
\end{equation}
where $z_i$ is the height in each ring $i$ of radius $R$ and $N_R$ is the number of particles in each ring.
Then, $A_1$ is given by equation (\ref{eq:A_1}):

\begin{equation}
A_1 = \sqrt{a_1^2+b_1^2}.
 \label{eq:A_1}
\end{equation}
The quantity $A_1$ defined in this way acts as a measure of the intensity of the bending mode \citep[e.g.][]{garcia-conde}. Figs.~\ref{fig:fig14} and \ref{fig:fig15} show $A_1$ for the perturbed models of barred and non-barred galaxies. The lighter, yellowish colors mark the areas where the warp is most intense. There is a predominance of these colors in the outermost radii of the disc in cases where the central galaxies are perturbed by the satellites of mass M2 and M3. Again, it can be noted the difference in the warp induced in the barred and non-barred galaxies, as well as the influence of the mass of the satellite on the intensity of this distortion. It is also possible to identify that the maximum intensity of the warp (very yellow regions) is reached after 10 Gyr. The greenish and light blue areas appearing in the inner regions of the disc ($0 < R < 20$ kpc) indicate that the asymmetry appears also in the central region, but with less intensity. The time it takes depends on the orbital radius of the satellite: when the satellite has a smaller orbit (R1), the distortions in the center and at the edges occur at the same time, at the start of the simulation. But for the satellite with larger radius (R3), the warp is first established on the edges of the disc and only after a few Gyr it also appears in the central regions.

We also examined the possibility of finding a correlation between the orientation of the warp and the orientation of the bar. The orientation of the bar $\theta_{\mathrm{bar}}$ is given by the equation~(\ref{eq:theta_b}):

\begin{equation}
 \mathrm{tan} \, \theta_{\mathrm{bar}} = \frac{b_2}{a_2}
 \label{eq:theta_b}
\end{equation}
with $a_2$ and $b_2$ defined by equations~(\ref{eq:a}) and (\ref{eq:b}). The warp orientation $\theta_{\mathrm{warp}}$ is calculated in the same way, but with $a_1$ and $b_1$ determined by the equations~(\ref{eq:a1}) and (\ref{eq:b1}). Our analysis found no correlation between $\theta_{\mathrm{bar}}$ and $\theta_{\mathrm{warp}}$, i.e. the maximum intensity of the warp is neither preferentially aligned with the bar nor perpendicular to it.

All the measurements of warp intensity indicated in which of the central galaxy models the warp signal is stronger. As these models differ in just two aspects -- the presence of gas and the presence of a bar, we needed to check if any changes in the warp were really due to the morphology of the galaxy or if the gas has some influence. Therefore we carried out a test, preparing a second version of the NB-M2R2 model, replacing all the gas particles with disc particles at the start of the simulation ($t=8$ Gyr). The mean height maps and the evolution of $W$ showed that the warp is still less intense in this case than in the barred galaxy, so it is not the presence of gas that inhibits vertical asymmetries. All the checks we made for this test are in the Appendix \ref{sec:appendix-a}.


\section{Discussion}
\label{sec:discussion}

In our simulations, the bar in the B models has undergone changes during the interactions. We used the $m = 2$ mode of the projected mass distribution to calculate the bar strength $A_2$ and found that the mass and orbital radius of the satellite determined the fate of the bar of the central galaxy. The satellites of mass M1 did not change the development of the bar; the satellites of mass M2 made it weaker and smaller; and the satellites of mass M3 destroyed it. These impacts on the bar occurred earlier in the case of fly-bys of satellites with orbital radius R1, later in interactions with satellites of orbital radius R2, and even later in minor mergers with satellites of orbital radius R3.

These results are consistent with the findings of \cite{ghosh}, which conclude that bar weakening is common in minor merger events, because it causes central mass enhancement, and \cite{athanassoula2003p}, which argues that the evolution of interacting galaxies depends strongly on the density of the satellite, leading to its destruction (if it has low mass) or to the bar destruction (if it is massive). \cite{zana} show that bars can be resilient and recover their strength after a fly-by with a satellite, however, in our simulations, there are several passages of the satellite. At the end of the orbit, in most models (except in M1R3 and M1R2), the perturber is absorbed into the disc of the central galaxy. Therefore, there is no recovery in the bar strength.

We used spherical satellites, but other results could have emerged if their shape had been different. \cite{ghosh} performed simulations with a barred central galaxy interacting with disc and spheroidal dwarf galaxies, and they conclude that spheroidal satellites are more impactful in weakening the bar because they are more efficient in the process of bringing stellar particles within the bar region. 

\cite{lokas2018} and \cite{peschken} indicate that the interaction with satellites can trigger bar formation or reinforce an existing bar, depending on the parameters of the interaction. Among other studies that analyzed the formation of tidally induced bars, \cite{purcell} investigated the influence of the Sagittarius dwarf galaxy on the formation of the spiral arms and bar of the Milky Way. But in their models, the satellite has two crossings with the Milky Way at distant radii from the center of the disc (30 and 15 kpc). In our simulations, the bar is weakened or destroyed when the satellites reach an orbital radius of $r \sim$ 5 kpc, being absorbed by the disc and directly impacting the region of the bar. Meanwhile, \cite{gauthier} investigated a M31 model that has bar formation after a few Gyr interacting with 100 satellites with great diversity of masses, initial orbital radii, orbit shapes and pericentric passages. The variety of configurations makes it difficult to make a direct comparison with our results. Despite this, it seems that interaction with satellites, depending on their properties, can lead to two scenarios: the formation of the bar or its weakening and destruction.

Both B and NB models developed warps induced by satellites. We quantified the intensity of the asymmetries by comparing mean height maps, analyzing the peaks in the azimuthal height profile ($W$) and measuring the intensity of the bending mode ($A_1$). $W$ captures a specific intensity for each instant of time. We found that the measurement is not extremely sensitive to the choice of radial ring. On the other hand, $A_1$ has the advantage of providing radial information at each instant of time. However, it ideally captures a dipole-type bending mode, but in our models, the warp has a spiral appearance for most of its evolution, so its orientation and intensity is not well defined at all instants.

Despite these restrictions, our study demonstrated that the characteristics of the satellite have a systematic impact on the induced warps. Satellites of mass M1 cause very small asymmetries in the central galaxies and satellites of masses M2 and M3 induce notable warps, but the ones with the greatest amplitude are the result of interaction with satellites of mass M3. The radial information provided by $A_1$ shows that the satellites of orbital radius R1 induce distortions in the center and edges of the disc at same time, in the beginning of the simulation; meanwhile, satellites of orbital radius R2 and R3 first distort the edges, and later, the asymmetries also appear in the center.

In \cite{Gomez2017} magneto-hydrodynamics simulations, warps have several causes besides interaction with satellites, and the galaxies interacted with more than one satellite. Despite the differences compared to this study, a common result is the influence of massive satellites in inducing more intense warps: in \cite{Gomez2017}, galaxies with strong vertical patterns interacted at least once with a satellite with mass $\sim 10^{10.5} \mathrm{M_{\odot}}$ during the last $\sim 5$ Gyr. \cite{grand} also concluded that massive satellites (masses $>10^{10} \mathrm{M_{\odot}}$) were relevant for vertical disc heating in their simulations of Milky Way-like galaxies. The inclination of the orbit of the satellite is also a factor that can impact the intensity of the induced warp. \cite{semczuk} found that the angle between the orbital angular momentum of the satellite and the angular momentum of the central galaxy that most preferentially lead to warp formation is $\sim 45$\textdegree. In our analysis, all orbits are polar, so we were not able to compare how the intensity of the warp varies in relation to this aspect. 

The main result of our analysis is that the induced warps are always stronger in the B models in comparison with the NB ones. This fact applies to all perturbed models and even to the isolated models: in agreement with \cite{chequers}, our isolated barred galaxy shows a subtle warp signal (at $t \sim 10$ Gyr), without any interaction. The difference between the B and NB models must not be due to the gas, since we did a test (appendix \ref{sec:appendix-a}) with one of the perturbed NB models replacing gas particles with stars and still the non-barred galaxy continued to suffer less warp. So it was not the presence of gas that was inhibiting intense warp. Another conclusion in common with \cite{chequers} is that bending waves should be long-lived features in disc galaxies. In our simulations, the induced warps last for several Gyr.


\section{Conclusions}
\label{sec:conclusions}

In this paper, we use $N$-body simulations to investigate the susceptibility of barred and non-barred galaxies to develop warps induced by satellites. Our models were composed of a central spiral galaxy (B and NB models) and spherical satellites with varying masses (M1, M2 and M3) and orbital radii (R1, R2 and R3). We analyzed how the interactions between the galaxies impact the development of the bar and the formation of warps. Our main findings are listed below:

\begin{itemize}
    \item In the interactions between satellites and the barred central galaxy, the bar can be preserved, weakened or destroyed. The level of bar weakening is determined by the mass of the satellite. And the time it takes depends on the initial orbital radius of the satellite.
    \item Warps in barred and non-barred galaxies are induced by the interaction with satellites. The mass of the perturber is the main influence on the intensity of the warp. These vertical asymmetries appear strongly at the edges and more weakly in the central regions. The initial orbital radius of the satellite determines when they will appear in the innermost part of the central galaxy.
    \item Warps are always stronger in the barred galaxies in comparison with the non-barred galaxies, in perturbed and isolated models.
\end{itemize}

In future studies, it would be interesting to explore models with satellite orbits of different inclinations and check their impact on the bar weakening and warp formation. The parameter space with galaxies with bars of different intensities, could also be investigated. Additional analysis of the vertical frequencies of the orbits may clarify the mechanism responsible for the response of a barred galaxy to external disturbances.


\section*{Acknowledgements}

AW acknowledges support from \textit{Coordenação de Aperfeiçoamento de Pessoal de Nível Superior - Brasil} (CAPES) – Finance Code 001. RM acknowledges support from the Brazilian agency \textit{Conselho Nacional de Desenvolvimento Cient\'ifico e Tecnol\'ogico} (CNPq) through grants 406908/2018-4 and 307205/2021-5, and from \textit{Funda\c c\~ao de Apoio \`a Ci\^encia, Tecnologia e Inova\c c\~ao do Paran\'a} through grant 18.148.096-3 -- NAPI \textit{Fen\^omenos Extremos do Universo}.

\section*{Data Availability}

The data supporting this article will be shared upon reasonable request to the corresponding author.


\bibliographystyle{mnras}
\bibliography{example}



\appendix

\section{Gas replacement experiment}
\label{sec:appendix-a}

\begin{figure}
    \includegraphics[width=\columnwidth]{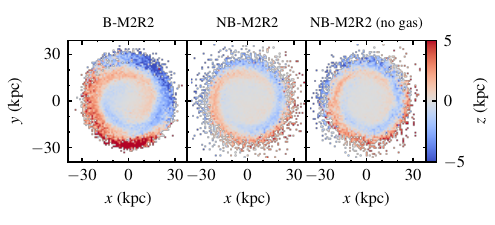}
	\caption{Maps of mean heights at $t=14$ Gyr comparing the models perturbed by a satellite of mass $0.5 \times 10^{10} \mathrm{M_{\odot}}$ and orbital radius $20$ kpc. From left to right there is the barred galaxy B-M2R2, the original non-barred galaxy NB-M2R2 and this same model but with the gas particles replaced by stars (NB-M2R2 (no gas)). B-M2R2 and NB-M2R2 models have already been shown in Fig.~\ref{fig:fig10}.}
	\label{fig:figA1}
\end{figure}

\begin{figure}
    \includegraphics[width=\columnwidth]{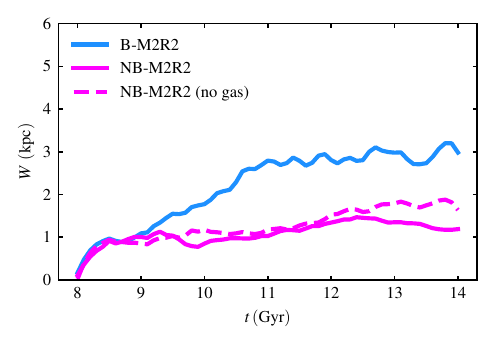}
	\caption{Evolution of maximum amplitude of warp $W$ for models perturbed by a satellite of mass $0.5 \times 10^{10} \mathrm{M_{\odot}}$ and orbital radius $20$ kpc. B-M2R2 and NB-M2R2 models have already been shown in Fig.~\ref{fig:fig12}.}
	\label{fig:figA2}
\end{figure}

In order to check if our conclusions about warps in B and NB models are reliable, we prepared a second version of the NB-M2R2 model, replacing the gas particles with stars at $t$ = 8 Gyr. This altered model has evolved up to 14 Gyr and we saved outputs every 0.05 Gyr, as we did with all the other simulations. Fig.~\ref{fig:figA1} shows the mean height maps comparing the original and the altered NB-M2R2 models with the B-M2R2 model. At the end of the simulation, in $t = 14$ Gyr, there is almost no difference between the NB models. To understand the evolution of warp in these models over time, we also analyzed the evolution of maximum amplitude of warp $W$, showed in Fig.~\ref{fig:figA2}. The warp intensity is similar in the gas-free model compared to the original, during the most part of the simulation, and the $W$ values are slightly lower at the end. Most importantly, both NB models have a lower warp intensity in comparison with the barred galaxy. These results indicate that it is not the presence of gas that inhibits vertical asymmetries in NB models. 

\bsp	
\label{lastpage}
\end{document}